\documentclass[aps,prb,twocolumn,showpacs,superscriptaddress,groupedaddress]{revtex4}  
\usepackage{graphicx}  
\usepackage{dcolumn}   
\usepackage{bm}        
\usepackage{amssymb}   
\usepackage{color}
\usepackage{amsmath}   
\usepackage{comment}
\usepackage{braket}

\begin{document}

\title{Coherent control of 40-THz optical phonons in diamond using femtosecond optical pulses}

\author{Tetsuya Kimata}
 \email{kimata.t.aa@m.titech.ac.jp}
 \altaffiliation[Also at: ]{Ground Systems Research Center, Ministry of Defense (ATLA), 2-9-54 Fuchinobe, Sagamihara 252-0206, Japan.}
 
\affiliation{ Laboratory for Materials and Structures, Institute for Innovative Research, Tokyo Institute of Technology, 4259 Nagatsuta, Yokohama 226-8503, Japan}
\affiliation{ Department of Materials Science and Engineering, Tokyo Institute of Technology, 4259 Nagatsuta, Yokohama 226-8503, Japan}

\author{Kazuma Yoda}
\affiliation{ Laboratory for Materials and Structures, Institute for Innovative Research, Tokyo Institute of Technology, 4259 Nagatsuta, Yokohama 226-8503, Japan}
\affiliation{ Department of Materials Science and Engineering, Tokyo Institute of Technology, 4259 Nagatsuta, Yokohama 226-8503, Japan}


\author{Hana Matsumoto}
\affiliation{ Laboratory for Materials and Structures, Institute for Innovative Research, Tokyo Institute of Technology, 4259 Nagatsuta, Yokohama 226-8503, Japan}
\affiliation{ Department of Materials Science and Engineering, Tokyo Institute of Technology, 4259 Nagatsuta, Yokohama 226-8503, Japan}

\author{Hiroyuki Tanabe}
\affiliation{ Laboratory for Materials and Structures, Institute for Innovative Research, Tokyo Institute of Technology, 4259 Nagatsuta, Yokohama 226-8503, Japan}
\affiliation{ Department of Materials Science and Engineering, Tokyo Institute of Technology, 4259 Nagatsuta, Yokohama 226-8503, Japan}

\author{Fujio Minami}
\affiliation{ Laboratory for Materials and Structures, Institute for Innovative Research, Tokyo Institute of Technology, 4259 Nagatsuta, Yokohama 226-8503, Japan}
\affiliation{ Department of Physics, Yokohama National University, 79-5 Tokiwadai, Yokohama 240-8501, Japan}


\author{Yosuke Kayanuma}
\affiliation{ Laboratory for Materials and Structures, Institute for Innovative Research, Tokyo Institute of Technology, 4259 Nagatsuta, Yokohama 226-8503, Japan}
\affiliation{ Graduate School of Sciences, Osaka Prefecture University, 1-1 Gakuen-cho, Sakai 599-8531, Japan}

\author{Kazutaka G. Nakamura}
 \email{nakamura.k.ai@m.titech.ac.jp}
\affiliation{ Laboratory for Materials and Structures, Institute for Innovative Research, Tokyo Institute of Technology, 4259 Nagatsuta, Yokohama 226-8503, Japan}
\affiliation{ Department of Materials Science and Engineering, Tokyo Institute of Technology, 4259 Nagatsuta, Yokohama 226-8503, Japan}

\date{\today}

\begin{abstract}
Coherent control is an optical technique to manipulate quantum states of matter. The coherent control of 40-THz optical phonons in diamond was demonstrated by using a pair of sub-10-fs optical pulses. The optical phonons were detected via transient transmittance using a pump and probe protocol.  The optical and phonon interferences were observed in the transient transmittance change and its behavior was well reproduced by quantum mechanical calculations with a simple model which consists of two electronic levels and shifted harmonic oscillators.

\end{abstract}

\maketitle

\section{Introduction}

Coherent control is a technique to manipulate quantum states in matter using optical pulses\cite{brumer,rice,scherer} and applied to electronic states, spins, molecular vibrations and phonons\cite{unanyan,weinacht,meshulach,omori1,katsuki,branderhorst,omori2,noguchi,higgins}.  The coherent control of optical phonons has been widely achieved in various materials such as semiconductors, semimetals, superconductors and dielectric materials\cite{nakamura3,Dekorsy1993,hase,cheng, Mashiko2018}. 
Recently, a theory of the coherent control of optical phonons by double-pulse excitation had been developed base on the simple quantum mechanical model with two-electronic bands and shifted harmonic oscillators\cite{nakamura1,nakamura2}.  We demonstrated the coherent control experiment on the optical phonons in a single crystal of diamond and analyzed the data by using the developed theory\cite{sasaki}.

The optical phonon in diamond has high frequency (approximately 40 THz) and is expected to be used as a qubit operating at room temperatures\cite{lee1,lee2,england1,england2}.  
The coherent optical phonons in diamond have been studied using femtosecond optical pulses and its generation mechanism has also been discussed with quantum mechanical calculations.\cite{nakamura1, sasaki, Ishioka2006, Zukerstein2018, Zukerstein2019, Yamada2019}
We have demonstrated the coherent control of amplitude and phase of the coherent optical-phonon oscillation in diamond using a pair of sub-10-fs infrared pulses at delays between  230 and 270 fs, in which two pump pulses were well separated\cite{sasaki}.  The observed behavior of the amplitude and phase were well reproduced by the developed theory.  
However, one of the paths,  in which electronic excitation and de-excitation occur in the different pump pulses,  was ignored in this delay range, because two pump pulses were well separated.  In the present work, we examined the coherent control of the optical phonons in diamond at pump-pump delays between -10 and 120 fs.     We analyzed the experiment using quantum mechanical calculations.

\section{Experimental}

\begin{figure}[htbp]
\includegraphics[width=8cm, bb=0 0 644 522]{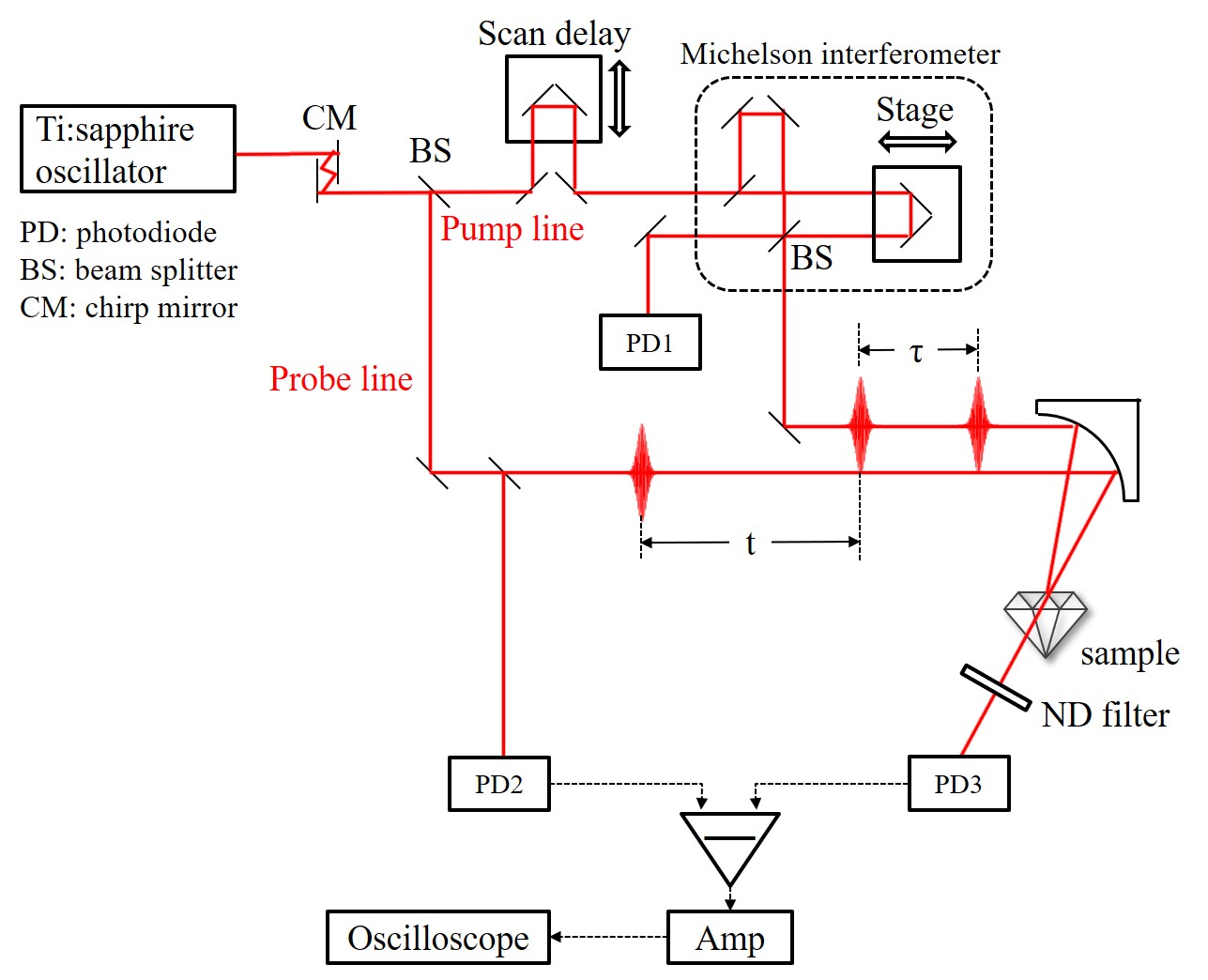}
\caption{Schematic of the experimental setup for coherent control measurements of optical phonon using the double pump-probe technique. This is a setup for the transient transmission measurement. Red and black dashed lines represent the optical path and the electric connection, respectively.}
\label{setup}
\end{figure}

\begin{figure}[htbp]
\includegraphics[width=7cm, bb=0 0 582 714]{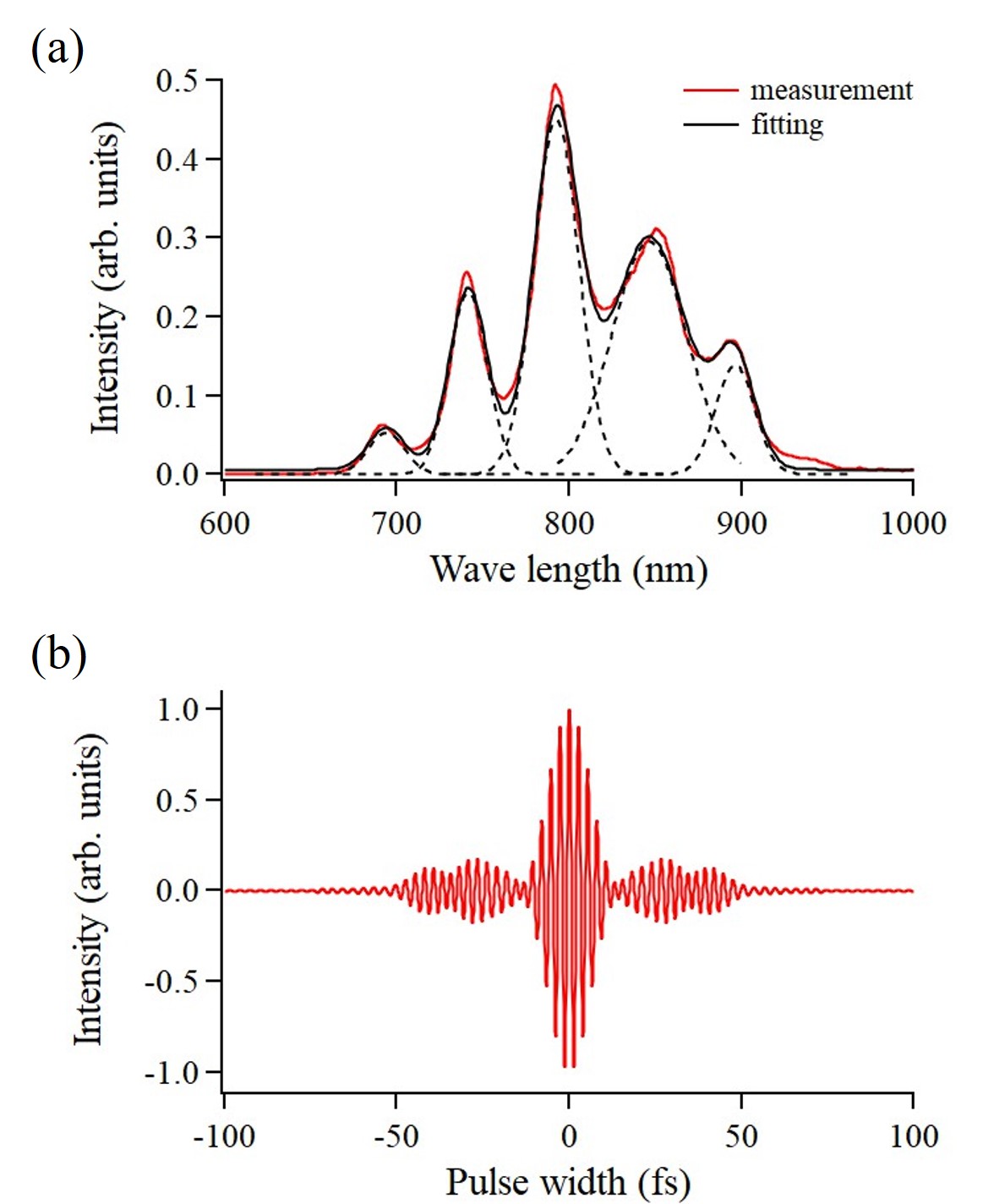}
\caption{(a) The measured spectrum of the ultrafast laser pulse (red) and the curve fitting with five Gaussian pulses (black). The dashed curves are each Gaussian pulses. (b) Electric field intensity generated from the five Gaussian curves.  Obtained parameters for five Gaussian pulses are listed in the text.}
\label{spectrum}
\end{figure}

The coherent optical phonons were investigated using a pump-probe-type transient-transmission measurement with two pump pulses (Fig. \ref{setup}). 
The used laser was a Ti:sapphire oscillator (FEMTOLASERS: Rainbow) operating with a repetition rate of 75 MHz.
Fig. \ref{spectrum} (a) shows the spectrum measured immediately behind the output port, which has peaks at 694, 742, 792, 846 and 896 nm, measured using a USB spectrometer (OceanOptics: USB2000). 
Fig. \ref{spectrum} (b) shows the pulse waveform generated by Fourier transformation of five Gaussian curves.
The pulse width was estimated to be 8.3 fs as full width at half maximum by using the frequency-resolved autocorrelation measurement (FEMTOLASERS: Femtometer).

The output from the Ti:sapphire oscillator was introduced to compensate the group-velocity dispersion using a pair of chirp mirrors and divided into two pulses by a beam splitter. One was used as a pump pulse and the other was used as a probe pulse.  The pump pulse was introduced to a scan-delay unit operating at 20 Hz in order to control a delay ($t$) between pump and probe pulses.  The pump pulse is introduced to  a home-made Michelson-type interferometer\cite{hayashi} to produce a pair of pump pulses (pump 1 and 2).
One optical arm of the interferometer was equipped with an automatic positioning stage (Sigma Tech Co. Ltd., FS-1050UPX), which moves with a minimum step of 1 nm.
The delay between pump 1 and 2, $\tau$ (fs), was controlled by the stage of Michelson interferometer in 0.5 fs steps. The optical interference of pump-pump delay was detected by a photodiode (PD1).

The probe pulse was picked up by a 95:5 beam splitter to measure the reference beam intensity at a photodiode (PD2). Thereafter, both pump and probe pulses were focused on the sample by using an off-axis parabolic mirror with a focal length of 50 mm. 
The transmitted pulse from the sample was detected with a photodiode (PD3).
By applying the opposite bias voltages to PD2 and PD3, we set the balanced detection before the experiment. 
Its differential signal, amplified with a low-noise current amplifier (Stanford Research Systems: SR570), was measured by a digital oscilloscope (Iwatsu: DS5534). To reduce the statistical error, the 32 000 signals were averaged and taken as the measured value. 
By converting the temporal motion of the scan delay unit to the pump-probe pulse duration, the temporal evolution of the transmittance change $\Delta T/T_0$ was obtained.
Here we used the heterodyne detection technique. The powers of the pump 1 and 2 and the probe were 21.2 mW, 21.3 mW, and 3.1 mW, respectively. 
The sample used was a single crystal of diamond with a [100] crystal plane, which was fabricated by chemical vapor deposition and obtained from EPD corporation. The type of diamond was intermediate between Ib and IIa and its size was 5 $\times$ 5 $\times$ 0.7 mm$^3$. The polarization of the pump and probe pulses were set along the [110] and [$-$110] axes, respectively.

\section{Results and discussion}

\subsection{Experimental results}

\begin{figure}[htbp]
\includegraphics[width=8cm, bb=0 0 547 453]{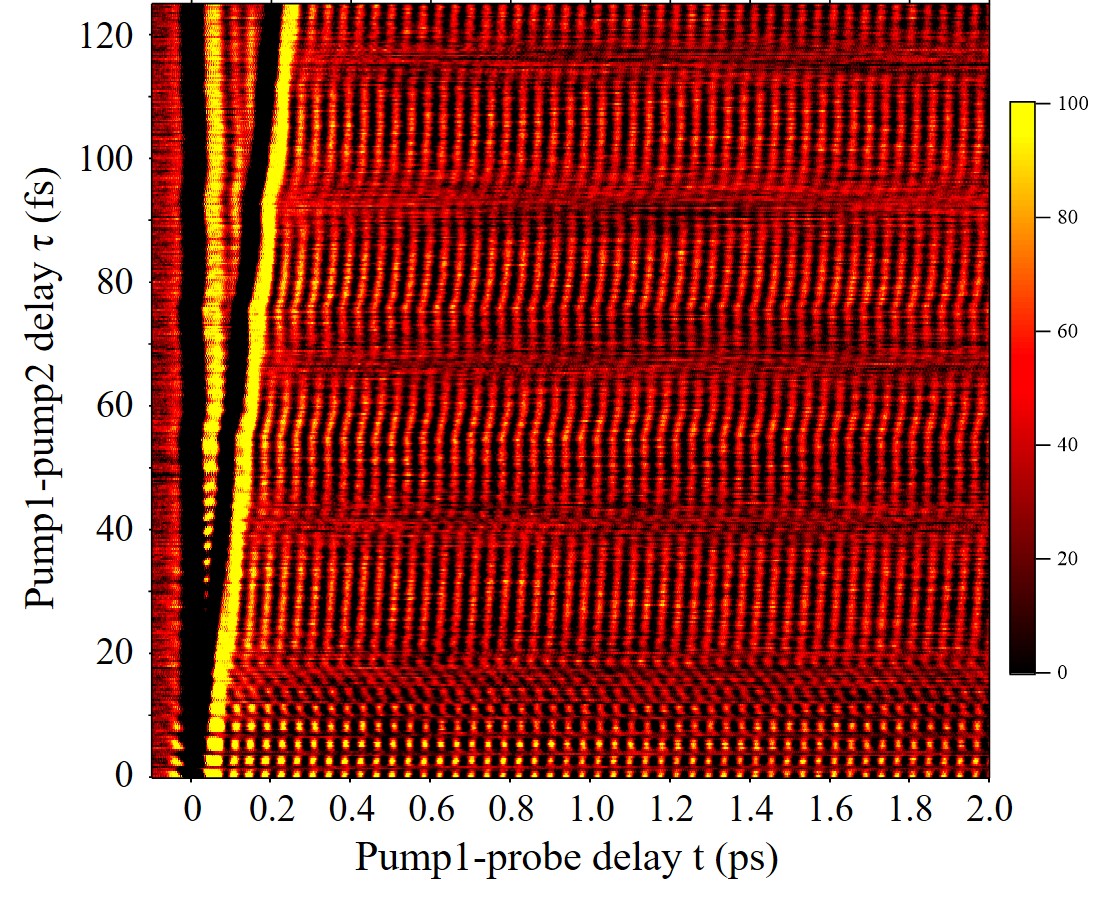}
\caption{Two-dimensional image map of the change in transition intensity with the pump1-probe delay (t) and pump1-pump2 delay ($\tau$).}
\label{2d}
\end{figure}

Figure \ref{2d} is a two-dimensional map of the transient transmittance change $\Delta T/T_0$ as a function of  the pump1-probe delay (t: horizontal axis) and pump1-pump2 delay ($\tau$: vertical axis). 
At the fixed pump1-pump2 delay ($\tau$), $\Delta T/T_0$ shows the sharp peak at delay zero between pump and probe and the successive oscillation with a frequency of 39.9$\pm$0.05 THz, which have been reported in a previous paper \cite{sasaki}.  The  oscillation is assigned to the optical phonons in diamond.
On the other hand, at a fixed pump-probe delay (t),  $\Delta T/T_0$ shows a rapid oscillation (approximately 380 THz) in addition to the oscillation due to the optical phonon (approximately 40 THz).  

\begin{figure}[htbp]
\includegraphics[width=8cm, bb=0 0 511 445]{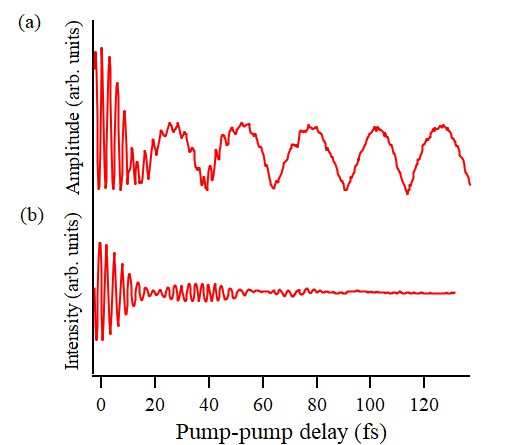}
\caption{The amplitude of (a) the controlled oscillation after pump 2 and (b) the optical interference against the pump-pump delay ($\tau$). The amplitude is normalized using that obtained after excitation after only pump 1; oscillation between the pump 1 and pump 2 irradiation timing.} 
\label{tau2}
\end{figure}

The Fourier transformation was performed between 0.5 ps $<$ 1.5 ps along the pump-probe delay after the irradiation of pump 2 at the each pump-pump delay.
The optical phonon amplitude was obtained by the integration of the Fourier-transformed data between 37.5 THz $\sim$ 42.5 THz, and then plotted against the pump-pump delay $\tau$ in Fig. \ref{tau2} (a).
Figure \ref{tau2} (a) shows slow and fast oscillations, with an oscillational period of approximately 25 and 2.7 fs, respectively.
The slow oscillation with a period of approximately 25 fs is observed at the pump-pump delay $\tau >$ 15 fs. This oscillation is consistent with the oscillating function in the range of the pump-pump delay between 230 and 270 fs in a previous study \cite{sasaki}. 
The amplitude is enhanced or suppressed at timing of the pump-pump delay which matches to integer or half-integer multiply of the phonon oscillation period. This is due to constructive or destructive interference of coherent phonons. 
On the other hand, the rapid oscillation with a period of approximately 2.7 fs is observed at the pump-pump delay $\tau <$ 15 fs. This oscillation continues on the oscillating function of the constructive and destructive interference of coherent phonons until approximately 80 fs. The rapid oscillation would be originated by the optical interference of the dual pulses because the influence of optical interference detected by PD1 (Fig. \ref{tau2} (b)) appeared in Fig. \ref{tau2} (a).

\subsection{Theoretical calculations}
In order to explain the behavior at the pump-pump delay $\tau <$ 50 fs, we performed the quantum mechanical calculations for the coherent control of optical phonons in a pulse-overlap region. 
The coherent optical phonons should be excited by the impulsive stimulated Raman scattering (ISRS) process \cite{nakamura2} at an off-resonant condition because the energy of the optical pulses (around 1.5 eV) was well below the direct band gap (7.3 eV) of diamond \cite{saslow,milden}.

We calculated the generation of the coherent phonons using a simple model with two-electronic levels and shifted harmonic oscillators.  
We used a model Hamiltonian of the electron-phonon system as 
\begin{eqnarray}
H &=& \hbar \omega b^\dagger b \ket{g}\bra{g} \nonumber\\
&+& \left( \epsilon + \hbar \omega b^\dagger b  +  \alpha \hbar \omega (b + b^\dagger)  \right) \ket{e}\bra{e},
\end{eqnarray}
where $\omega$ is the phonon frequency, $\epsilon$ is the excitation energy, and $\alpha$ represents the electron-phonon coupling.  $\ket{g}$ and $\ket{e}$ denote the electronic ground and excited states, respectively.  $b^\dagger$ and $b$ denote creation and annihilation operators for the phonon.  The interaction Hamiltonian with the optical pulse is given by
\begin{eqnarray}
H_I = \mu E(t) \left( \ket{e}\bra{g} + \ket{g}\bra{e}  \right),
\end{eqnarray}
where $\mu$ is the transition dipole moment and $E(t)$ is the electric field of the pulse.
We used a density operator formalism and the second-order perturbation, which is described in previous papers \cite{nakamura2, Nakamura2019}.

\begin{figure}[htbp]
\includegraphics[width=7cm, bb=0 0 570 630]{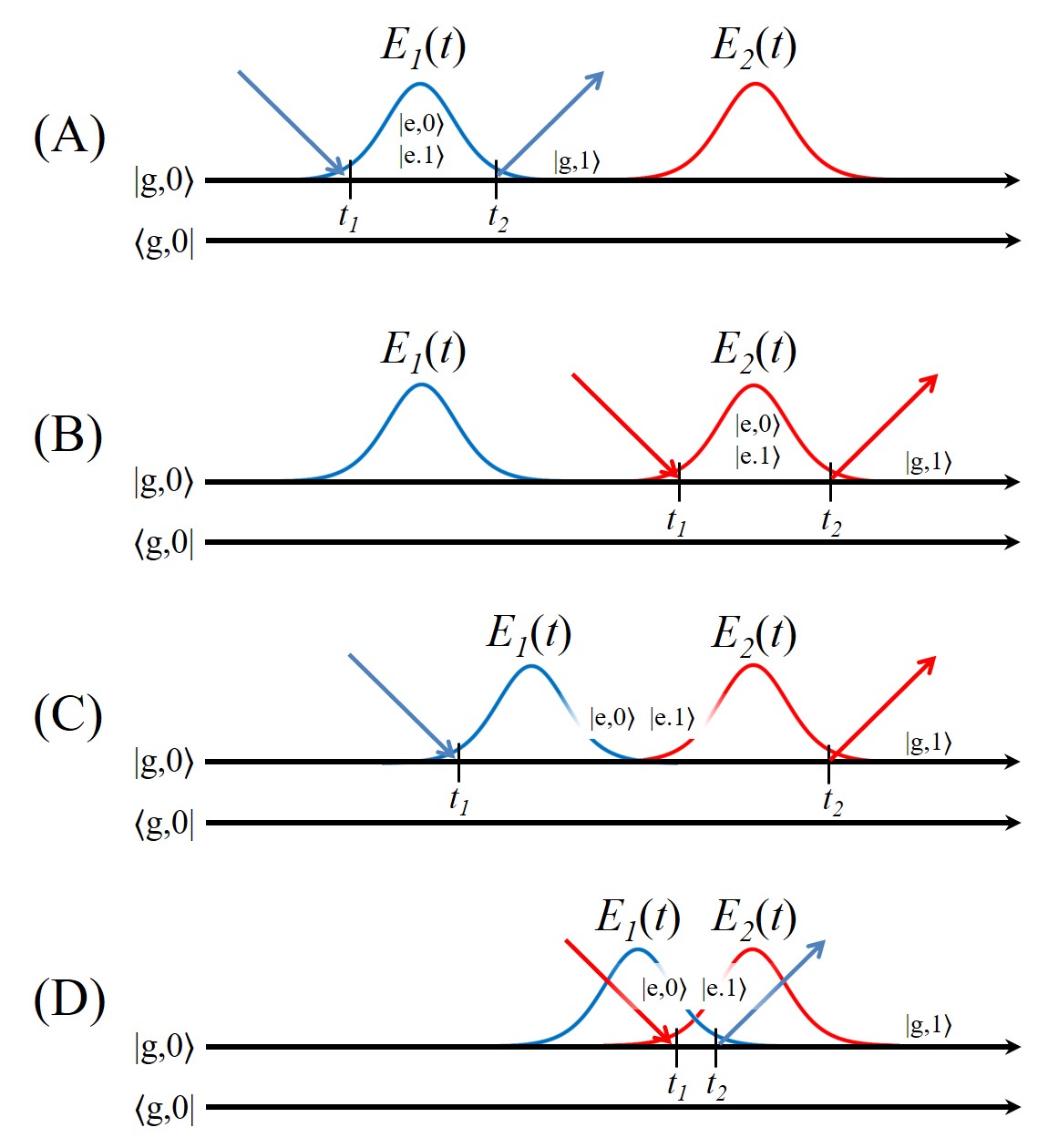}
\caption{Diagram of the ISRS paths. Transition occurs by only the pump 1 (A) or the pump 2 (B). (C) and (D) represent the transition occurring by both the pump 1 and pump 2. Time flows from the left to the right. The upper part of the diagram shows an envelope of the optical pulse. Interaction between the optical pulse and the system occurs at time $t_1$ and $t_2$.}
\label{path}
\end{figure}

Figure \ref{path} shows the double-side Feynman diagrams for ISRS paths on the double pulse excitation \cite{nakamura3}.
 $E_1(t)$ and $E_2(t)$ are the electric field of pump 1 and pump 2.
Interaction between the optical pulse and the system occurs at time $t_1$ and $t_2$. The first diagram (Fig. \ref{path}(A)) indicates the path A, in which the phonons generated by only the pump 1; the electronic excitation and de-excitation occur within the pulse. 
In a similar way, the phonons are generated by the only pump 2, which is the path B and shown in the second diagram (Fig. \ref{path}(B)).  Figures \ref{path}(C) and (D) show the path C and D in which the phonons are excited by both pump 1 and 2. The phonon-excitation and de-excitation are induced by the pump 1 and pump2, respectively, in the path C and vis versa in the path D.

By using the second-order perturbation, the density operator for all paths, $\rho(t)$, is obtained as
\begin{align}
\rho(t)=&\alpha\left(\frac{\mu}{\hbar}\right)^2e^{-i\omega t}\sum_{j=1}^2\sum_{k=1}^2\int_{-\infty}^{t}dt_2\int_{-\infty}^{t_2}dt_1E_j(t_1)E_k(t_2)\notag\\
&\times e^{-\frac{i}{\hbar}\epsilon(t_2-t_1)}\left(e^{i\omega t_1}-e^{i\omega t_2}\right)|g, 1\rangle\langle g, 0|\notag\\
&+H.c.,
\end{align}
where $\ket{g,1}$ and $\bra{g,0}$ indicate the state vector for the electronic ground state with one-phonon state and that with zero-phonon state, respectively.  It should be noted that time $t$ should be enough after the irradiation of pump 2.
Details for the calculation of the density operator are described elsewhere \cite{nakamura3, Nakamura2019}.

In the calculation, we used the optical pulse composed of five Gaussian pulses: 
\begin{align}
E_1(t) = \sum_{k=1}^5 E_k \exp\left(-\frac{t^2}{\sigma_k^2}\right) \cos\left(2\pi\cdot\left(\Omega_k+\theta t \right)\cdot t\right),
\end{align}
where $E_k$, $\Omega_k$, and $\sigma_k$ are the ratio of electric-field strengths, the optical frequency, and the pulse width of each component. The parameters were determined by curve fitting of the measured spectrum with five Gaussian-shape spectrum.  The fitted spectrum is shown in Fig. 2 (a).  The obtained parameters are ($E_k, \Omega_k [1/$fs], $\sigma_k [1/$fs]) =(0.045, 0.432, 0.035),(0.197, 0.405, 0.032), (0.385, 0.379, 0.033), (0.253, 0.354, 0.048), and (0.119, 0.335, 0.023). 
$\theta$ is the linear chirp rating. 
The electric field of pump 2 is defined as $E_2(t) = E_1(t-\tau)$, where $\tau$ is a delay between pump 1 and pump2. 
$\theta$ is set to $5 \times 10^{-4}$ (fs$^{-2}$) by comparing the first order optical interference derived from the electric field of the two pump pluses with the detected optical interference (Fig. \ref{tau2}(b)).

The mean value of the phonon coordinate is obtained by $\braket{Q(t)} = {\rm Tr} \{Q \rho (t) \}$, where $Q \equiv \sqrt{\hbar /2 \omega} (b + b^\dagger) $ and Tr indicates that the trace should be taken over the electronic and phonon variables.

\begin{figure}[htbp]
\includegraphics[width=7cm, bb=0 0 526 555]{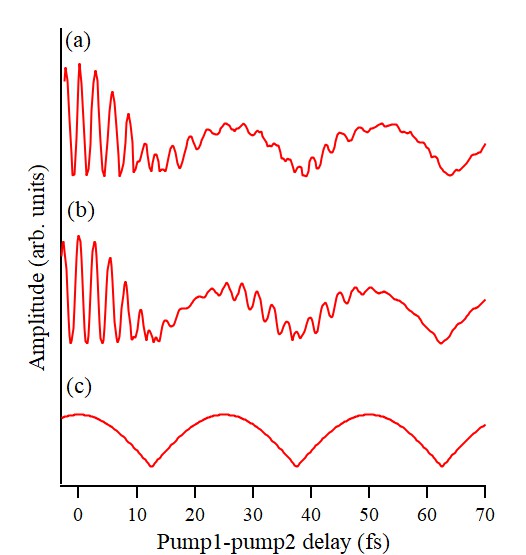}
\caption{(a) The measured amplitude of the controlled oscillation after pump 2 on the pulse overlap region. (b) The calculated phonon amplitude for all paths. (c) The calculated phonon amplitude for path A and path B. }
\label{tau}
\end{figure}

The calculated phonon amplitude for all paths shown in Fig. \ref{tau}(b) represents well the experimental data (Fig. \ref{tau}(a)).
The rapid oscillation with a period of approximately 2.7 fs at the pump-pump delay $\tau < 15$ fs is caused by the optical interference, which is  represented by paths C  and D.  This interference pattern corresponds well to the optical interference (shown in Fig. 4 (b)) and no electronic-coherence effect, which was reported in the coherent-control experiments on GaAs at resonance conditions \cite{Nakamura2019}, has not been observed. 
On the other hand, the slow oscillation with a period of approximately 25 fs is caused by the phonon interference, which is due to the path A and B.  In fact, the calculated phonon amplitude via path A and B shows only the slow oscillation as shown in Fig. \ref{tau} (c). As already reported \cite{sasaki}, the phonon amplitude after the second pump pulse irradiation is a sum of two sinusoidal functions induced by each pulse.

\section*{ACKNOWLEDGMENTS}
The authors thank Prof. Y. Shikano of Keio University and Riho Tanaka of Tokyo Tech for their efforts in the early stage of this work. 
This work was supported in part by JSPS KAKENHI Grant Numbers 15K13377, 17K19051, 17H02797, 19K03696 and 19K22141, Collaborative Research Project of Laboratory for Materials and Structures, Institute of Innovative Research, Tokyo Institute of Technology.


\begin{thebibliography}{99}
\bibitem{brumer} P. Brumer and M. Shapiro, Chem. Phys. Lett. {\bf 126}, 541 (1986).

\bibitem{rice}
S. A. Rice, D. J. Tannor, R. Kosloff, J. Chem. Soc., Faraday Trans. {\bf 82}, 2423 (1986).

\bibitem{scherer}
N. F. Scherer, R. J. Carlson, A. Matro, M. Du, A. J. Ruggiero, V. Romero-Rochin, J. A. Cina, G. R. Fleming, S. A. Rice, J. Chem. Phys. {\bf 95}, 1487 (1991).

\bibitem{unanyan}
R. Unanyan, M. Fleischhauer, B. W. Shore, K. Bergmann, Opt. Commun. {\bf 155}, 144 (1998).

\bibitem{weinacht}
T. C. Weinacht, J. Ahn, P. H. Bucksbaum, Nature {\bf 397}, 233 (1999).

\bibitem{meshulach}
D. Meshulach, Y. Silberberg, Nature {\bf 396}, 239 (1998).

\bibitem{omori1}
K. Ohmori, H. Katsuki, H. Chiba, M. Honda, Y. Hagihara, K. Fujiwara, Y. Sato, K. Ueda, Phys. Rev. Lett. {\bf 96}, 093002 (2006).

\bibitem{katsuki}
H. Katsuki, H. Chiba, B. Girard, C. Meier, K. Ohmori, Science {\bf 311}, 1589 (2006).

\bibitem{branderhorst}
M. P. A. Branderhorst, P. Londero, P. Wasylczyk, C. Brif, R. L. Kosut, H. Rabitz, I. A. Walmsley, Science {\bf 320}, 638 (2008)

\bibitem{omori2}
K. Ohmori, Annu. Rev. Phys. Chem. {\bf 60}, 487 (2009)

\bibitem{noguchi}
A. Noguchi, Y. Shikano, K. Toyoda, S. Urabe, Nat. Commun. {\bf 5}, 3868 (2014).

\bibitem{higgins}
G. Higgins, F. Pokorny, C. Zhang, Q. Bodart, M. Hennrich, Phys. Rev. Lett. {\bf 119}, 220501 (2017).

\bibitem{nakamura3}
K. G. Nakamura, K. Yokota, Y. Okuda, R. Kase, T. Kitashima, Y. Mishima, Y. Shikano, Y. Kayanuma, Phys. Rev. B {\bf 99}, 180301(R) (2019).

\bibitem{Dekorsy1993} T. Dekorsy, W. K\"utt, T. Pfeifer, and H. Kurz, Europhys. Lett. \textbf{23}, 223 (1993).

\bibitem{hase}
M. Hase, K. Mizoguchi, H. Harima, S. Nakashima, M. Tani, K. Sakai, M. Hangyo, Appl. Phys. Lett. {\bf 69}, 2474 (1996).

\bibitem{cheng}
Y.-H. Cheng, F. Y. Gao, S. W. Teitelbaum, K. A. Nelson, Phys. Rev. B {\bf 96}, 134302 (2017).

\bibitem{Mashiko2018} H. Mashiko, Y. Chisuga, I. Katayama, K. Oguri, H. Masuda, J. Takeda, and H. Gotoh, Nat. Commun. \textbf{9}, 1468 (2018). 

\bibitem{nakamura2} K. G. Nakamura, Y. Shikano, Y. Kayanuma, Phys. Rev. B {\bf 92}, 144304 (2015).

\bibitem{nakamura1} K. G. Nakamura, K. Ohya, H. Takahashi, T. Tsuruta, H. Sasaki, S. Uozumi, K. Norimatsu, M. Kitajima, Y. Shikano, Y. Kayanuma, Phys. Rev. B {\bf 94}, 024303 (2016).

\bibitem{sasaki} H. Sasaki, R. Tanaka, Y. Okano, F. Minami, Y. Kayanuma, Y. Shikano, K. G. Nakamura, Sci. Rep. {\bf 8}, 9609 (2018).

\bibitem{lee1} K. C. Lee, M. R. Sprague, B. J. Sussman, J. Nunn, N. K. Langford, X.-M. Jin, T. Champion, P. Michelberger, K. F. Reim, D. England, D. Jaksch, I. A. Walmsley, Science {\bf 334}, 1253 (2011).

\bibitem{lee2} K. C. Lee,  B. J. Sussman, M. R. Sprague, P. Michelberger, K. F. Reim, J. Nunn, N. K. Langford, P. J. Bustard, D. Jaksch, I. A. Walmsley, Nat. Photon. {\bf 6}, 41 (2012).

\bibitem{england1} D. G. England, P. J. Bustard, J. Nunn, R. Lausten, B. J. Sussman, Phys. Rev. Lett. {\bf 111}, 243601 (2013).

\bibitem{england2} D. G. England, K. A. G. Fisher, J.-P. W. MacLean, P. J. Bustard, R. Lausten, K. J. Resch, B. J. Sussman, Phs. Rev. Lett. {\bf 114}, 053620 (2015).

\bibitem{Ishioka2006} K. Ishioka, M. Hase, M. Kitajima, H. Petek, Appl. Phys. Lett. {\bf 92}, 231916 (2006).

\bibitem{Zukerstein2018} M. Zukerstein, M. Koz\'{a}k, F. Troj\'{a}nek, P. Mal\'{y}, Diamond \& Related Matter. {\bf 90}, 202-206 (2018).

\bibitem{Zukerstein2019} M. Zukerstein,  F. Troj\'{a}nek, B. Rezek, Z. \v{S}ob\'{a}\v{n}, M. Koz\'{a}k, P. Mal\'{y}, Appl. Phys. Lett. {\bf 115}, 161104 (2019).

\bibitem{Yamada2019} A. Yamada, K. Yabana, Phys. Rev. B {\bf 99}, 245103 (2019). 


\bibitem{hayashi} S. Hayashi, K. Kato, K. Norimatsu, M. Hada, Y. Kayanuma, K. G. Nakamura, Sci. Rep. {\bf 4}, 4456 (2014).

\bibitem{saslow} W. Saslow, T. K. Bergstresser, M. L. Cohen, Phys. Rev. Lett. {\bf 16}, 354 (1996).

\bibitem{milden} R. P.  Milden. In {\sl Optical engineering of diamonds}, edited by R. P. Milden and J. R. Rabeau, (Wiley-VCH, Weinheim, 2013), pp. 1-34.

\bibitem{Nakamura2019} Kazutaka Nakamura. In {\sl Quantum Phononics}. Springer Tracts in Modern Physics {\bf 282} (Springer Nature, 2019).











\end{thebibliography}
\end{document}